\documentclass[useAMS,usegraphicx]{mn2e}
\usepackage{rotate}
\usepackage{times}
\newif\ifAMStwofonts
\AMStwofontstrue

\def\gs{\mathrel{\hbox{\rlap{\hbox{\lower4pt\hbox{$\sim$}}}\hbox{$>$}}}}
\def\ls{\mathrel{\hbox{\rlap{\hbox{\lower4pt\hbox{$\sim$}}}\hbox{$<$}}}}

\def\Msun{M$_{\odot}$}

\def\chandra{{\it Chandra}}
\def\astroe{{\it Astro-E~2}}

\def\xmm{{\it XMM-Newton}}

\def\et{{et al.\ }}

\def\ugc{{UGC~3973}}
\def\3c{{3C~273}}

\def\rg{{\thinspace r_{\rm g}}}
\def\ka{{K$\alpha$}}

\def\nixxviii{{Ni~\textsc{xxviii}}}

\def\nh{{N_{\rm H}}}

\def\cm{{\rm\thinspace cm}}
\def\erg{{\rm\thinspace erg}}
\def\eV{{\rm\thinspace eV}}

\def\ga{{\rm\thinspace gauss}}

\def\keV{{\rm\thinspace keV}}

\def\Msun{\hbox{$\rm\thinspace M_{\odot}$}}

\def\s{{\rm\thinspace s}}
\def\ks{{\rm\thinspace ks}}

\def\ph{{\rm\thinspace photons}}

\def\phpkevpscmps{\hbox{$\ph\keV^{-1}\cm^{-2}\s^{-1}\,$}}

\def\ergpscmps{\hbox{$\erg\cm^{-2}\s^{-1}\,$}}

\def\ergps{\hbox{$\erg\s^{-1}\,$}}

\def\pscm{\hbox{$\cm^{-2}\,$}}

\title[The possible 8~keV emission feature in \ugc]
      {
A possible line-like emission feature at 8~keV in the Seyfert 1.2 \ugc  
      }

\author[L. C. Gallo et al.]
       {L. C. Gallo,$^1$ 
	A. C. Fabian,$^2$ 
	Th. Boller,$^1$
	and W. Pietsch$^1$ \\
$^{1}$ Max-Planck-Institut f\"ur extraterrestrische Physik, Postfach 1312, 85741 Garching, Germany \\
$^{2}$ Institute of Astronomy, University of Cambridge, Madingley Road, Cambridge CB3 0HA\\
}
\date{Accepted. Received. }
\pagerange{\pageref{firstpage}--\pageref{lastpage}}
\pubyear{2005}
\begin{document}
\maketitle
\label{firstpage}

\begin{abstract}
Two short X-ray exposures ($< 3600\s$ each) of the radio-quiet Seyfert 1.2 
galaxy UGC~3973 (Mrk~79) were conducted with \xmm\ as part of
an AGN snap-shot survey.  
In this paper we concentrate 
on the significance and possible origin of a narrow $8\keV$ (rest frame)
line-like emission feature detected with the pn instrument during the 
second observation.
Simulations show that the feature is significant at $96.0-98.4\%$ confidence,
depending on what a priori assumptions are made.
The feature cannot be attributed to background contamination  
and appears to be 
variable (or transient) since it was not detected in the
first observation of \ugc\ six months earlier.  However, a constant feature
cannot be completely dismissed, based on the $90\%$ upper-limit on 
the flux from the first observation.
There is some indication
that the feature is variable over the duration of the second 
observation as well.
We discuss various models (e.g. Ni emission, recombination edges, outflows,
disc lines) which could potentially produce an emission feature 
at such energies.

\end{abstract}



\section{Introduction}
\label{sect:intro}

\ugc\ (Mrk~79; $z = 0.022$) is an optically well-studied, radio-quiet 
(e.g. Ho 2002), Seyfert 1.2 galaxy (e.g. Pietsch \et 1998).
Numerous black hole mass estimates exist (e.g. Wandel \et 1999;
Vestergaard 2002; Peterson \et 2004) which yield a value of
$\sim 6 \times 10^{7}\Msun$.

\ugc\ was observed with \xmm\ (Jansen et al. 2001) at two epochs
separated by six months as part of an AGN snap-shot survey, the results of
which are presented in Gallo \et (2005; hereafter G05).  These short 
observations
constituted the first analysis of the X-ray spectrum of \ugc\ above $2\keV$.
In addition to an iron emission line at $\sim 6.4\keV$, which was detected at 
both epochs, G05 report the detection of a narrow emission feature 
at $7.99\pm0.06\keV$ 
(rest frame) during the second (low-flux) observation.
In this paper we elaborate the discussion of this $\sim 8\keV$ feature.  
We have scrutinised the reliability of the detection and discuss physical 
scenarios which could manifest a line-like feature at such energies.

\section{Observations and data reduction}
\label{sect:data}
During both observations of \ugc\ all three EPIC instruments functioned
normally.  The pn (Str\"uder et al. 2001) and MOS (MOS1 and MOS2; 
Turner \et 2001) cameras were operated in small window mode 
with the medium filter in place. 
A summary of the observations is provided in Table~\ref{tab:log}.
Details of the data reduction (which were conducted with 
{\tt XMM-SAS v6.1.0}) are given in G05.
\begin{table}
\begin{center}
\caption{Log of \xmm\ observations of \ugc. 
The date and \xmm\ revolution number when the observation was conducted are 
given in columns (1) and (2), respectively.
The total amount of useful exposure (GTI) is shown in column (4), and
the estimated number of $0.3-10\keV$ source counts is reported in column (5).  
}
\begin{tabular}{ccccc}                
\hline
(1) & (2) & (3) & (4)  & (5)  \\
Date  &  Rev. & Instrument & Exposure & Counts \\
year.mm.dd & & & (s) & ($0.3-10\keV$) \\
\hline
2000.10.09 & 153  &  pn    &  1680 & 21114 \\
           &      &  MOS1  &  1861 & 5519  \\
           &      &  MOS2  &  1838 & 5600  \\
\hline
2001.04.26 & 253  &  pn    &  3590 & 20521 \\
           &      &  MOS1  &  5737 & 9442  \\
           &      &  MOS2  &  5736 & 9553  \\
\hline
\label{tab:log}
\end{tabular}
\end{center}
\end{table}


\section{Spectral analysis}
\label{sect:fit}

\begin{table*}
\begin{center}
\caption{The broadband ($0.3-10\keV$) fits to the pn data of \ugc\ at each
epoch found by G05. 
The revolution number is given in column (1).  
In column (2) the fit quality ($\chi^2_\nu$/dof) is stated.
Columns (3)-(13) are fit parameters and quantities which are measured 
from the model.
Column (3) shows the measured column density at the redshift
of \ugc\ ($10^{20}\pscm$).  
In column (4) the blackbody temperature is given and in column (5) the
power-law photon index.  The parameters of an absorption edge: energy and 
optical depth are given in columns (6) and (7), respectively.
The emission line parameters: energy, width, and equivalent 
width, as measured with a Gaussian profile are given in columns
(8), (9), and (10), respectively. 
The observed $0.3-10\keV$ flux corrected for Galactic absorption
in units of $10^{-12}\ergpscmps$
is reported in column (11).  
The rest frame $2-10\keV$ luminosity in units of $10^{43}\ergps$ is given in
column (12).  
In column (13) the luminosity ratio between the blackbody and power-law component
(corrected for Galactic and intrinsic
absorption) in the $0.3-10\keV$ band is estimated.
Values marked with an $f$ denote fixed parameters.
}
\begin{tabular}{ccccccccccccc}                
\hline
(1) & (2) & (3) & (4)  & (5) & (6) & (7) & (8) & (9) & (10) & (11) & (12) & (13)\\
 Rev. &  $\chi^2_\nu$/dof & $\nh_{z}$  &  $kT$ &  $\Gamma$  & $E_{edge}$ & $\tau$ & $E_{line}$ & $\sigma$ & $EW$ & $F_{0.3-10}$ & $L_{2-10}$ & $\frac{L_{bb}}{L_{po}}$   \\
  &  &  &  (eV) &  & (eV) & & (keV) & (eV) & (eV) &  & &   \\
\hline
153 & $0.91/480$ & $<0.63$ & $122\pm7$ & $1.85\pm0.04$ & -- & -- & $6.33^{+0.08}_{-0.09}$ & $1^f$ & $117\pm15$ & $42.75$ & $2.18$ & $0.13$ \\
\hline
253 & $1.00/502$ & $<0.99$ & $116\pm7$ & $1.67\pm0.04$ & $726\pm15$ & $0.46^{+0.12}_{-0.08}$ & $6.45\pm0.07$ & $122^{+68}_{-65}$ & $216^{+42}_{-27}$ & $21.78$ & $1.28$ & $0.20$ \\
 &  &  & & & & & $7.99\pm0.06$ & $1^{f}$ & $161\pm21$ &  &  &  \\
\hline
\label{tab:cont}
\end{tabular}
\end{center}
\end{table*}
\begin{figure*}
\begin{minipage}[]{0.45\hsize}
\scalebox{0.32}{\includegraphics[angle=270]{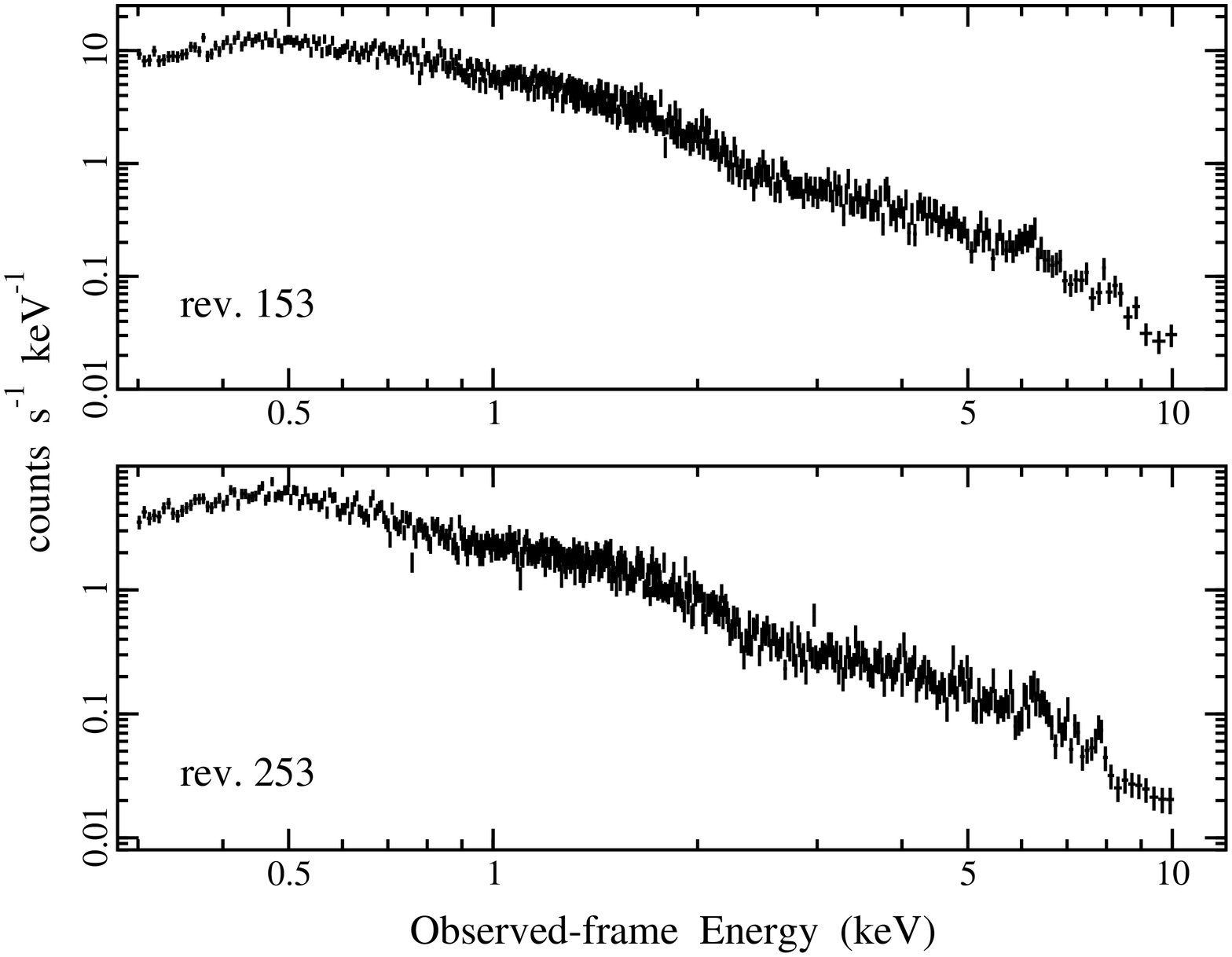}}
\end{minipage}
\hfill
\begin{minipage}[]{0.45\hsize}
\scalebox{0.32}{\includegraphics[angle=270]{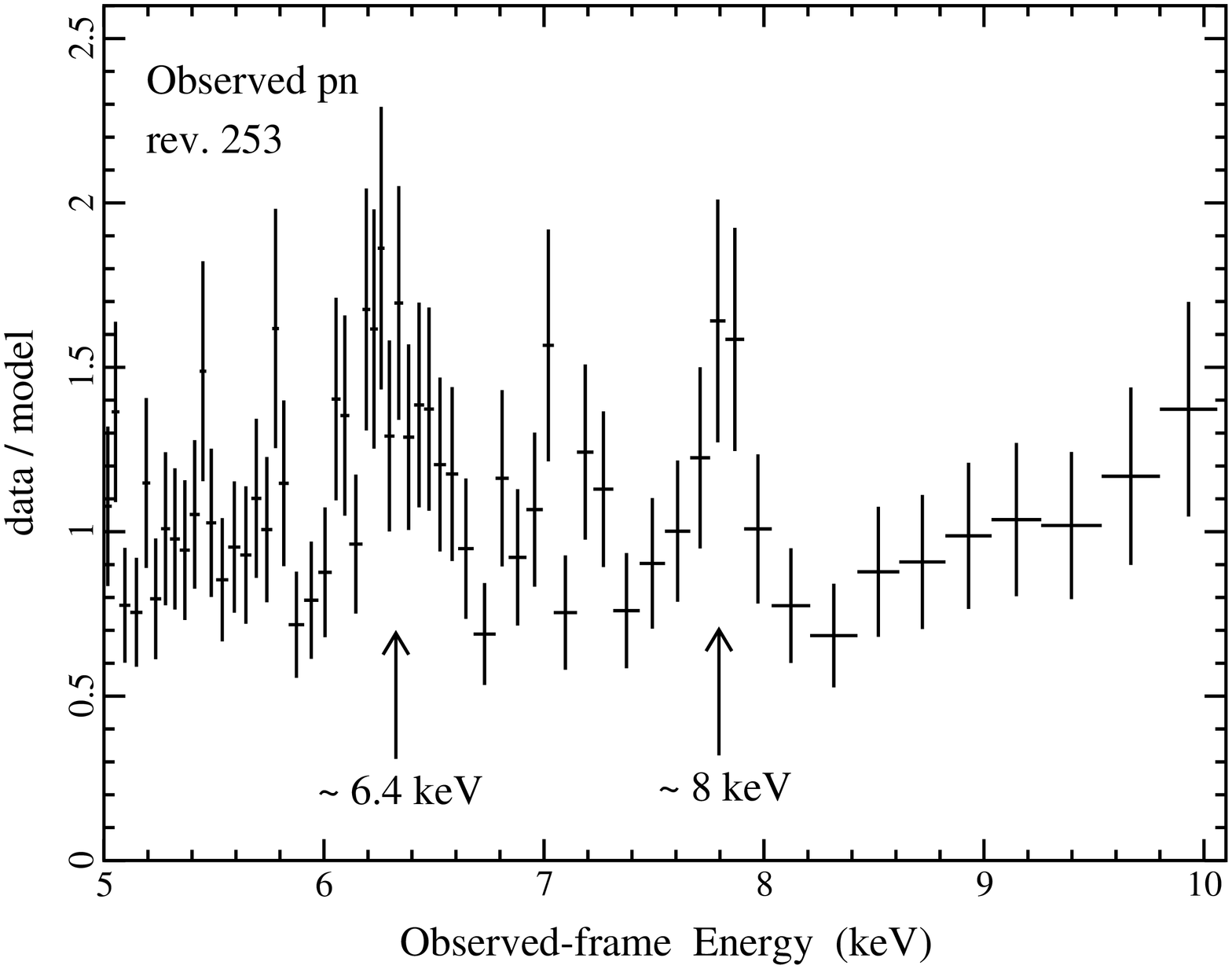}}
\end{minipage}
\hfill
\caption
{\label{fig:po}
On the left-hand side are the observed $0.3-10\keV$ count spectra of \ugc\
during revolution 153 (top panel) and revolution 253 (bottom panel).
\ugc\ was about twice as bright during the first observation.
On the right-hand side are the residuals (data/model) remaining in the 
$5-10\keV$ range after fitting the broadband spectrum from revolution 153 
with a blackbody plus power-law and absorption edge as described in 
Table~\ref{tab:cont}.  Excess residuals at $\sim 6.2\keV$ and
$\sim 7.8\keV$ in the observed frame correspond to $\sim 6.4\keV$ and 
$\sim 8\keV$ in the
source frame, respectively.
}
\end{figure*}

The spectra were grouped such that each bin contained at least 20
counts. Spectral fitting was performed using {\tt XSPEC v11.3.1} (Arnaud
1996).
Fit parameters are reported in the rest frame of the object, although most
of the figures remain in the observed frame.
The quoted errors on the model parameters correspond to a 90\% confidence
level for one interesting parameter (i.e. $\Delta\chi^2$ = 2.7 criterion).
K-corrected luminosities were derived assuming isotropic emission.
A value for the Hubble constant of $H_0$=$\rm 70\ km\ s^{-1}\ Mpc^{-1}$ and
a standard cosmology with $\Omega_{M}$ = 0.3 and $\Omega_\Lambda$ = 0.7
were adopted.
A value for the Galactic column density toward \ugc\ of 
$5.69 \times 10^{20}\pscm$ (Dickey \& Lockman 1990) was assumed throughout.

\subsection{A $7.99\keV$ feature}
\label{sect:eight}

A discussion of the broadband ($0.3-10\keV$) and apparent spectral features
was presented in G05.  In this analysis we assume the best-fitting
broadband phenomenological model (Table~\ref{tab:cont})
introduced by G05, and concentrate on the high-energy spectrum.
The MOS data are analysed for consistency, but in the interest of brevity
they are discussed only when they appear discordant with the pn data.

During revolution 253 (the second observation) \ugc\ was about half as bright
as it was at the first epoch, and at high-energies it exhibited a
flatter spectrum.  In addition to neutral (or slightly ionised) iron emission,
which was present at both epochs, during revolution 253 a convincing 
emission-like feature was detected in the rest frame at $7.99 \pm 0.06\keV$
($\sim 7.82\keV$ in the observed pn data; Figure~\ref{fig:po}).
When fitted with a Gaussian profile the feature was 
intrinsically narrow ($\sigma$ was fixed at $1\eV$), and had an equivalent 
width of $EW = 161\pm21 \eV$ and flux of $f = (1.74\pm0.95) \times 10^{-13}
\ergpscmps$.
The addition of the Gaussian profile improved the 
existing broadband model by $\Delta\chi^2 = 8.1$ for 2 additional 
parameters ($502$ dof).
Allowing the width of the Gaussian profile to vary did not improve the fit,
nor was the measured value much different than the assumed value 
of $\sigma = 1\eV$.

Attempts to incorporate the $8\keV$ and the $\sim 6.4\keV$ features 
into a single broad line-like profile were unsuccessful.  
The single broad feature
resulted in a poorer fit and large residuals between $6-8\keV$. 

\subsection{Robustness of the detection}
\label{sect:robust}

\subsubsection{Significance of a physically motivated feature 
in the $7-10\keV$ range}

Given that the feature appears to be transient and
that the physical process which created it is unknown, estimating the
significance of it is challenging.
There are processes which, in principle, could produce emission lines
in the $7-10\keV$ range, but it is uncertain if they are at work here.
As such, we estimate the significance of the feature twice.  In this section 
we assume there is a priori physical motivation to find a line between
$7-10\keV$ (see Section~\ref{sect:diss}).  In the next section
we take a more conservative approach and consider the probability of finding
such a feature in any spectral channel.

In order to estimate the statistical significance we conducted Monte
Carlo simulations similar to the analysis outlined by Porquet \et (2004).
Under the assumption of the broadband spectrum presented in Table~\ref{tab:cont}
(without the Gaussian profile at $7.99\keV$), we simulated 1000 pn spectra
mimicking the photon statistics expected from a $3600\s$ exposure.
The faked spectra were grouped and fitted in the same manner as the spectrum
of \ugc.  To this model a Gaussian profile with a fixed width of $\sigma = 1\eV$
was added.  The two additional
free parameters were the line energy, which was limited between $7-10\keV$, and
the line normalisation.  Doing this to the source spectrum resulted in 
$\Delta\chi^2 = 8.1$. In this manner the $\Delta\chi^2$ was estimated for
each of the 1000 fake spectra.  In addition, the initial line energy was
stepped in $50\eV$ increments 
between $7-10\keV$ and the fit was restarted.  
The highest $\Delta\chi^2$ was recorded.
This step was taken in order to properly sample (oversample) the pn energy
resolution, which is $\sim 150\eV$ at $\sim 7\keV$.

A value of $\Delta\chi^2 > 8.1$ was obtained in 16 of the 1000 fake spectra, implying 
a detection significance of approximately $98.4\%$ confidence for the feature
in \ugc.  

\subsubsection{The ``true'' significance} 

We need to consider the possibility that $\Delta\chi^2 = 8.1$ can be
exceeded by a spurious event occuring in any observed energy channel.
However, we limit the energy range investigated to $2.3-6.4\keV$ and
$7.0-10.0\keV$.  The logic in ignoring the spectrum below $2.3\keV$ is
to avoid the calibration uncertainties and astronomical effects
(e.g. warm-absorbers in the source or Galaxy) which would affect the 
observations.  We also neglect the $6.4-7.0\keV$ range, where an emission
feature would be expected.

Following the procedure outlined in the previous section, we find that
$\Delta\chi^2 = 8.1$ is exceeded in 20 out of 1000 trials, resulting in a
confidence level of $98.0\%$.  Moreover, the overall significance of
the observed feature must account for the total number of observations
performed (in this case 2).
Therefore, by making no assumptions of the physical origin of the feature,
the emission-like line at $8\keV$ is detected with 
approximately $96.0\%$ confidence.

\subsection{Sanity checks}
\label{sect:check}

\subsubsection{Background contamination and instrumental emission lines}

\begin{figure}
\rotatebox{270}
{\scalebox{0.32}{\includegraphics{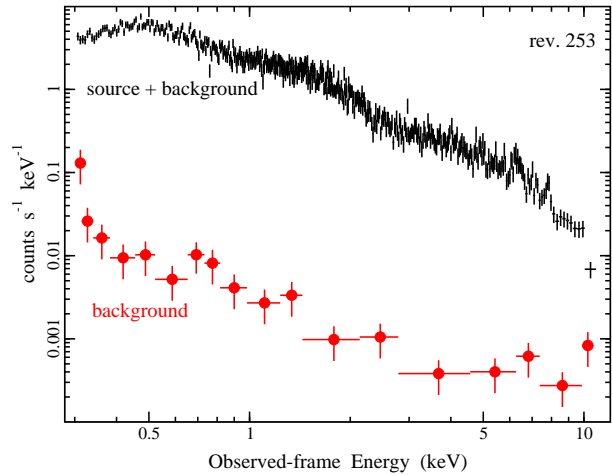}}}
\caption{The top curve of data points (black crosses) represent the 
source plus background pn spectrum during observation 253.
The lower curve (red dots) is the background pn spectrum extracted from
a source-free region and scaled to the source plus background extraction
region.
}
\label{fig:bg}
\end{figure}

As the observations were conducted in small window mode and with short
exposures the background level is very low.
During both observations, \ugc\ was significantly detected above the 
background level (Figure~\ref{fig:bg}).
The total number of source plus background counts collected by the pn
instrument in the $7-10\keV$ range during revolution 253 was 413.  In
comparison, only 8 background counts were collected in the $7-10\keV$ range
from an extraction region of the same size as the source cell.

To understand the background emission produced by the pn detector Freyberg
\et (2004) analysed $313\ks$ of data obtained in closed filter position
during the first three years of operations.
In doing so they accurately determined the spatial and temporal behaviour
of background spectral features.  The background features most relevant to
our analysis are the emission lines from Ni~\ka\ ($E = 7.48\keV$) and Cu~\ka\ 
($E = 8.05\keV$).  These lines originate from the electronics board mounted
below the CCD and they are spatial inhomogeneous across the camera.
Emission is weakest in the centre of the detector (close to the 
on-axis position) which is located above the venting hole in the circuit
board (see figure 4 of Freyberg \et).  When the pn detector is operated in
nominal small window mode, as was the case for both observations of \ugc,
the exposed part of the CCD is located above the venting hole and contamination
from Ni and Cu is minimised.

\begin{figure}
\rotatebox{270}
{\scalebox{0.36}{\includegraphics{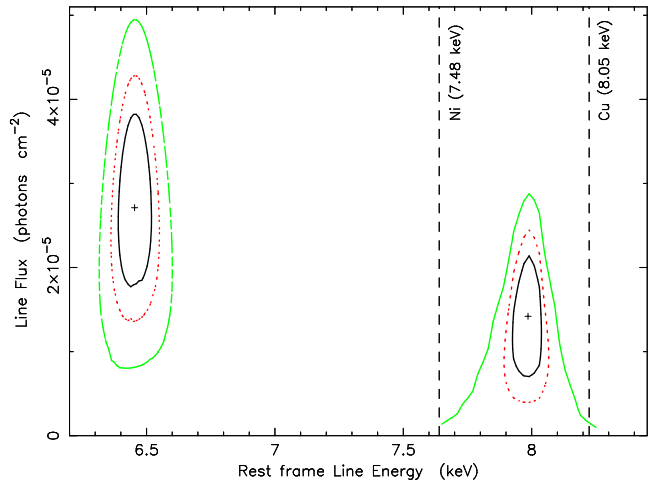}}}
\caption{The plot show the 1, 2, and $3 \sigma$ confidence levels 
(line flux as a function of rest frame line energy) from fitting a Gaussian
profile at $\sim 6.45\keV$ and another at $\sim 7.99\keV$.
The vertical dashed lines mark the energies (shifted to the source frame)
of the background emission lines (Cu and Ni) in the region around 
the $\sim 7.99\keV$ feature.
}
\label{fig:sigs}
\end{figure}

Furthermore, and more importantly, the $8\keV$ (rest frame) emission line
is found at $\sim 7.8\keV$ in the observed frame.  The energy is formally
inconsistent with any background emission features at greater than $3\sigma$
(Figure~\ref{fig:sigs}).  The combination of the low background level, observed
energy of the feature, and the minimal contamination expected from background
lines in small window mode, makes it very unlikely that the feature 
could arise from instrumental effects.

\subsubsection{MOS spectra}

\begin{figure}
\rotatebox{270}
{\scalebox{0.32}{\includegraphics{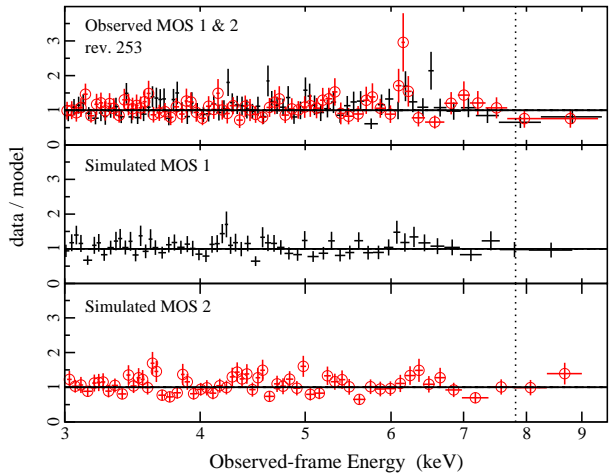}}}
\caption{In the top panel are the observed MOS 1 (black crosses) and 
MOS 2 (red open circles) residuals (data/model) remaining after fitting
the $2.5-10\keV$ data from revolution 253 with a power-law.  
There is no detection of a
$\sim 8\keV$ feature (marked by the dotted vertical line) as seen in the pn.
The residuals shown in the middle and lower panels result from fitting a 
power-law to simulations of the MOS observations (from rev. 253) which
include an $\sim 8\keV$ feature as found in the pn.  
There are no indications of an $\sim 8\keV$ emission line
in the simulated observations.
The lack of a MOS detection appears to be a sensitivity issue.
}
\label{fig:mossim}
\end{figure}

The broadband spectral modelling of the MOS data are in good agreement
with the pn results; however the $\sim 8\keV$ feature is not detected.
Simulations were conducted to examine if the null detection could be a
consequence of the lower sensitivity of the MOS cameras at $8\keV$
compared to the pn.
To the best fitting model obtained with the MOS data a Gaussian profile
was added with the parameters (energy, width, and normalisation) as measured
from the pn data.  The MOS observations from revolution 253 were then 
simulated.  The residuals resulting from fitting a power-law to the
$2.5-10\keV$ simulated spectra are shown in Figure~\ref{fig:mossim}.
An emission line is not detected in the simulated MOS observations
indicating that the absence of an $8\keV$ feature in the MOS is most
likely a sensitivity issue.

\subsection{Is the feature transient?}
\subsubsection{Long-term variability}

The question as to whether the $8\keV$ feature is variable or transient
is of fundamental importance in determining its origin.
Approximately six months earlier during the first observation (rev. 153)
there was no significant detection of a feature around $8\keV$.  
When fixing a Gaussian profile at $8\keV$ with the same parameters as 
found during revolutions 253, we determine the $90\%$ upper-limit on the
line flux to be $1.57 \times 10^{-13}\ergpscmps$, which is still 
consistent with the $90\%$ confidence levels reported for the flux of the
$8\keV$ feature during revolution 253 (Section~\ref{sect:eight}).

The continuum flux of \ugc\ during revolution 153 was twice as high 
compared to revolution 253, and this could possibly explain the absence
of a firm detection of the line during the first observation.
\begin{figure}
\rotatebox{270}
{\scalebox{0.32}{\includegraphics{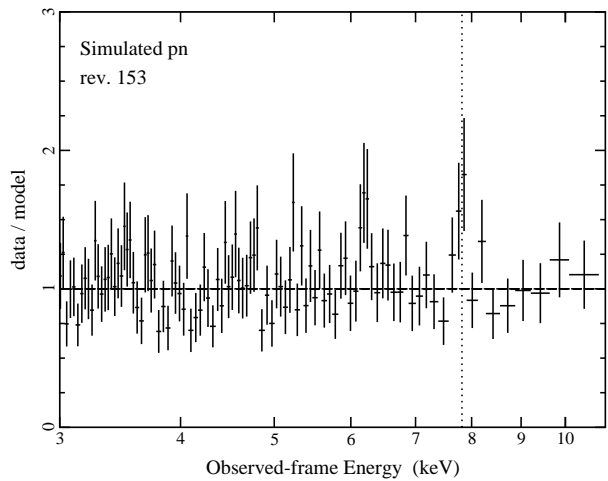}}}
\caption{A simulation of the pn spectrum from revolution 153 including
an $\sim 8\keV$ emission line with the same flux as measured during 
revolution 253 (marked by the dotted vertical line)
Assuming the line flux remained constant, the emission line
should have been detected during revolution 153 despite the continuum being
twice as bright.
}
\label{fig:pnsim}
\end{figure}
To clarify the situation, a simulation of the pn observation during revolution
153 was carried out.  To the best-fit broadband model measured during
revolution 153 a Gaussian profile was included, with the same parameters as
measured during revolution 253.  In addition to the 
higher continuum flux, the high-energy spectrum is also steeper during
revolution 153.  Consequently, a {\it constant-flux} emission line should
have been detectable with about the same strength and significance as
found at the later epoch (Figure~\ref{fig:pnsim}).  
The addition of a Gaussian profile with three free parameters (energy,
width, and normalisation) to the broadband simulated spectrum improves
the fit by $\Delta\chi^2 = 8.6$.

This suggests that the $8\keV$ feature may be variable or
transient, but we can not completely exclude the possibility of a constant-flux 
emission line.

\subsubsection{Rapid variability}

\begin{figure}
\rotatebox{270}
{\scalebox{0.32}{\includegraphics{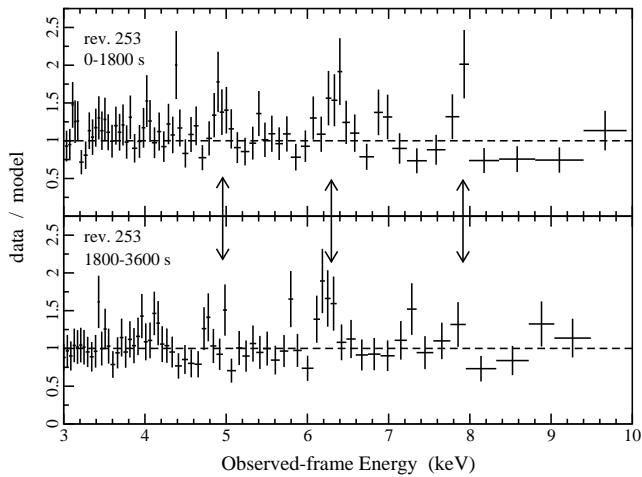}}}
\caption{The ratio resulting from fitting a power-law to the 
$2.5-10\keV$ pn band during two different intervals of revolution 253.  
The spectrum from the first half of the 
observation ($< 1800\s$) is shown in the top panel.  In the lower panel,
the spectrum from the second half ($> 1800\s$) of the observation is displayed.
Three double-headed arrows mark the possible line 
emission at
about 5, 6.4, and 8\keV.  All three features are visible in the first half
of the observation, but only the $\sim 6.4\keV$ line is present in the 
second half.
}
\label{fig:int}
\end{figure}

Due to the very low background during the observation it is possible to examine
short-term variability in the spectrum.  Over the course of the $\sim 3600\s$
exposure (revolution 253) light curves of \ugc\ are consistent with a 
constant and there is no change in the broadband continuum flux.
In order to examine possible variability in the features themselves, the 
spectrum from revolution 253 was divided into two halves each containing 
about $1800\s$ of data.  
The $2.5-10\keV$ spectra from both intervals were then fitted with a 
power-law, the ratios of which are shown in Figure~\ref{fig:int}.

\begin{table*}
\begin{center}
\caption{High-energy spectral features at two $1800\s$ intervals during 
revolution 253.
Column (1): interval exposure.  Column (2) is the power-law normalisation
in units of $10^{-3}\phpkevpscmps$ at $1\keV$.  
Columns (3) shows the rest frame of the line feature.  Column (4)
marks the fit improvement achieved by adding the component to the fit from the
row above.  Columns (5), (6), and (7) are the width, equivalent width and 
flux of each respective line feature.
Values marked with an $f$ denote fixed parameters.
}
\begin{tabular}{ccccccc}                
\hline
(1) & (2) & (3) & (4)  & (5) & (6) & (7) \\
Time  & $n$  & $E$ & $\Delta\chi^{2}$ & $\sigma$ & $EW$ & $f_l$ \\
Interval (\s) & & (\keV) & & (\eV) & (\eV) & ($10^{-13}\ergpscmps$) \\
\hline
$0-1800$ & $2.48^{+0.47}_{-0.39}$  & $6.48^{+0.07}_{-0.10}$ & & $< 220$ & $229^{+189}_{-131}$ & $2.87^{+2.21}_{-1.55}$ \\
  &  & $8.05\pm0.06$ & $5.3/2$ & $1^{f}$ & $212^{+113}_{-93}$ & $2.37^{+1.69}_{-1.51}$ \\
  &  & $5.03\pm0.06$ & $9.6/2$ & $1^{f}$ & $117^{+64}_{-66}$ & $1.72^{+0.83}_{-0.99}$ \\
\hline
$1800-3600$ & $2.54^{+0.58}_{-0.40}$  & $6.38^{+0.07}_{-0.06}$ & & $< 130$ & $198^{+164}_{-98}$ & $2.44^{+1.42}_{-1.21}$ \\
  &  & $8^{f}$ & $1.7/1$ & $1^{f}$ & $97^{+83}_{-71}$ & $< 2.39$ \\
  &  & $5^{f}$ & $0.2/1$ & $1^{f}$ & $9^{+38}_{-9}$ & $< 0.86$\\
\hline
\label{tab:inter}
\end{tabular}
\end{center}
\end{table*}

During both intervals the $\sim 6.4\keV$ line is present with approximately
the same strength and flux (Table~\ref{tab:inter}). Interestingly,
the $\sim 8\keV$ feature is significantly detected only in the first interval.
Moreover, there is another narrow spectral feature at $\sim 5\keV$,
which is also only detected during the first half of the observation. 

During the second interval only the $\sim 6.4\keV$ emission line is 
statistically required.  If we fix a Gaussian profile at $5\keV$ and another at
$8\keV$ and fit the data from the second interval, we notice that both 
lines have 
decreased in strength by about the same amount ($\Delta EW \approx -110\eV$).
Meanwhile, the strength and flux of the $\sim 6.4\keV$ emission line,
and the continuum flux and photon index remain constant throughout the
observation (Table~\ref{tab:inter}).
However, the changes in the $5$ and $8\keV$ features are trends as opposed
to significant variations since, within $90\%$ confidence levels, only the
equivalent width of the $5\keV$ feature is different.

These possible variations suggest that if both line-like features are real 
they may be 
related in such a way that they vary simultaneously.
Since the features are only observed for the first half of the observation
it is impossible to estimate how long they were present in the spectrum prior
to the onset of the observation.
The high-energy spectral fit parameters 
for each of the two intervals are shown in Table~\ref{tab:inter}.

\section{Discussion}
\label{sect:diss}

There are several reports of emission lines at energies
inconsistent with iron emission.  These features tend to be redshifted
to energies between $5-6\keV$ (e.g. Turner \et 2002, 2004; Porquet \et 2004),
and reports of blueshifted emission features are few.
Extreme blueshifts, on the order of $\sim 0.7c$ have been reported in 
two objects (Yaqoob \et 1999; Wang \et 2003).  Both these sources are 
radio-loud, thus a mechanism to launch material at such velocities can be 
envisaged.  Another blueshifted line was reported at $7.6\keV$ in a
radio-quiet, narrow-line Seyfert 1 galaxy (Ghosh \et 2004).
There are some possible physical explanation as to why there are more 
reports of redshifted features than blueshifted ones, though it could simply be
that the sensitivity of current X-ray instruments (i.e. \xmm\ and \chandra)
drops quickly at energies above $\sim 7\keV$.

\subsection{Atomic transitions}

\subsubsection{Emission lines}
From the {\tt CHIANTI v4.2} database (Dere \et 1997; Young \et 2003), we find
that an atomic transition arising around $8\keV$ originates from 
H-like Ni ($E = 8.0730\keV$).
In the time-averaged spectrum (reported in Table~\ref{tab:cont})
the best-fit line energy is formally inconsistent with Ni at more than 
$3\sigma$.  However, in the first half of the split spectrum 
(Table~\ref{tab:inter}) the measured line energy is comparable with
\nixxviii.  This raises a second problem in that an extraordinary abundance
of nickel is required to produce such a strong line.  Assuming solar
abundances for iron (Anders \& Grevesse 1989), we estimate that a minimum 
nickel overabundance 
of $\sim 15\times$ solar is necessary to account for
the average flux of the $\sim 8\keV$ feature over
the duration of the observation.

Gastaldello \& Molendi (2004) suggested an overabundance of nickel, probably
due to enrichment from SN~Ia (Dupke \& Arnaud 2001), to explain the excess
flux between $7-8\keV$ in the spectrum of the Perseus cluster.
This scenario does not seem valid for \ugc\ for a number of reasons.
Firstly, the overabundance is quite extreme ($> 15\times$ solar).  Secondly,
the $\sim 8\keV$ in \ugc\ appears to be variable on various time scales.
Flux variations would not be noticed from supernova emission
unless there was bulk motion of ejecta, which is at odds with the narrowness
of the measured feature.  
Of course, we cannot completely rule out that the environment was enriched
at an earlier epoch, and now the variability is due to accretion processes.
Thirdly, though not a necessary condition, a Ni
overabundance does not explain the possible detection of the variable 
$\sim 5\keV$ emission line.

\subsubsection{Radiative recombination edges}
Radiative recombination edge and continuum (RRC) of ionised iron
can result in line-like features above $7\keV$, specifically $9.28\keV$
and $8.83\keV$ for H- and He-like iron, respectively. 
The width of the feature reveals the characteristic electron
temperature ($kT_e$).  For a photoionised plasma the electron temperature
can be significantly less than the ionisation potential; thus the RRC feature
can appear quite narrow (see reviews by Liedahl 1999; Kahn \et 2002).
Radiative recombination features have been proposed to describe the 
X-ray afterglow spectra of some gamma-ray bursts 
(e.g. Yoshida \et 2002; Piro \et 2000).

The $8\keV$ feature in \ugc\ can be well fitted with a RRC
($\chi^2_\nu$/dof $= 1.00/501$) with
a plasma temperature of $kT_e < 140\eV$.  The observed threshold energy
(taking into consideration the host-galaxy redshift)
would require an outflowing velocity of approximately $0.14$ or $0.10~c$,
depending on the ionisation state of iron.   The RRC should also
be accompanied by a redshifted, ionised, emission line.  Unfortunately,
the redshifted emission line, regardless of the ionisation, should appear
close to, or within, the neutral Fe~\ka\ feature observed between $6-6.5\keV$;
thus it is not possible to derive any meaningful limits on a  
redshifted, ionised line component.

\subsection{Relativistic disc line and orbiting hot spots}

\begin{figure}
\rotatebox{270}
{\scalebox{0.32}{\includegraphics{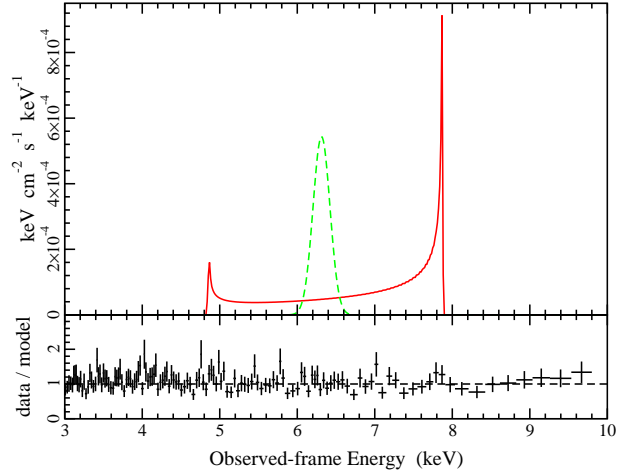}}}
\caption{Upper panel: 
The emission line profiles used to fit the near-neutral iron
emission and the narrow feature at $8\keV$.  
The red solid line is the disc line profile as described in the text,
and the green dashed line is the Gaussian profile used to fit the 
$\sim 6.4\keV$ emission line.
The
red horn of the disc line profile is at about $4.9\keV$ which could
describe the residuals seen at that energy in the top panel of 
Figure~\ref{fig:int}.
Lower panel: The residuals (data/model) remaining in the $3-10\keV$ band
after applying a power-law plus the two line profiles shown in the upper
panel to the pn data of \ugc\ during rev. 253.
}
\label{fig:discline}
\end{figure}

There are a number of examples (e.g. Della Ceca \et 2005, and references
within) where narrow emission features have been attributed to the ``horns'' of
the disc line profile (Fabian \et 1989).  
Our attempt to model the feature with a disc line profile resulted in a good
fit ($\chi^2_\nu$/dof $= 1.01/502$; Figure~\ref{fig:discline}).  
The energy of the disc line was fixed
to $6.7\keV$ as allowing it to remain free did not enhance the fit.
The equivalent width was $EW \approx 400\eV$.
Emission was assumed to be coming from a thin ring between $15-15.5\rg$
($\rg = GM/c^2$).
The derived disc inclination ($i = 60^{+6}_{-3}$~degrees) is consistent with
the optical classification of \ugc\ as a Seyfert 1.2 (Pietsch \et 1998).
The horns of the disc line profile are observed at $\sim 7.8$ and $4.9\keV$
(observed frame), which accurately correspond to the energies of the
$8\keV$ feature under investigation, as well as the slight excess at 
$\sim 5\keV$ seen in Figure~\ref{fig:int}.


Assuming a $\sim 6 \times 10^{7}\Msun$ black hole, 
implies that the radiation only
needs to flash briefly ($15\rg/c \approx 4500\s$).  This is on the same
order as the lower-limit established from the apparent variability of the 
lines in Figure~\ref{fig:int}.  The effect could be of radiation
propagating from the centre shining on an irregularity in disc height, perhaps
an inflowing lump or some sort of ridge in the disc.

Localised hot spots which illuminate
a small portion of the disc (e.g. Dov\v ciak \et 2004) could also produce
such a narrow feature at $8\keV$.
The possible $5\keV$ line seen in Figure~\ref{fig:int} appears as though 
it could be displaying similar variability as the $8\keV$ line.  If we
insist on connecting the behaviour of these two features
(which we do not), then it would be
challenging to describe both of them as arising under the same
physical conditions in the hot spot model.

\subsection{Outflows}

Both red and blueshifted $absorption$ features have been reported in
radio-quiet AGN (e.g. Nandra \et 1999; Matt \et 2005; Dadina \et 2005),
and the aborted jet scenario (Ghisellini \et 2004) works well at describing 
them.

In principle the aborted jet scenario can
also describe emission features, but this would imply that the line is 
being emitted at a large distance from the central region.  
If attributed to blueshifted iron, the $\sim 8\keV$ emission line in \ugc\
requires a outflowing velocity between $0.15-0.25~c$ (depending on the 
ionisation state of iron).
The high velocity measured in the X-rays is inconsistent with velocities 
seen at other wavelengths.

If we once again suppose that the $8\keV$ and $5\keV$ features are 
interconnected we can speculate that both are originating from slightly
ionised iron ($E = 6.5\keV$), and moving in opposite directions, but with
the same speed ($\approx 0.2~c$).  The slightly weaker $5\keV$ line would
be consistent with suppressed emission from matter moving away from the 
observer.  Such a condition could arise in a jet outflow, for example from
material which is ejected in both directions at the same time.  
Such a twin-jet scenario involving emission lines 
is not unprecedented and has been proposed for the 
Galactic binary SS~433 (e.g. Fabian \& Rees 1979; see 
Brinkmann \et 2005 for recent X-ray spectra).

\section{Conclusions}

We presented the analysis of an $8\keV$ emission line feature 
in the Seyfert 1.2, \ugc.  
We have intensely scrutinised the detection and determine that the
feature is significant at a confidence level between $96.0-98.4\%$, depending
on what a priori assumptions are made. 
The fact that the feature is blue shifted with respect to iron emission is intriguing.
However, since the exposure was so short ($< 3600\s$) it is not possible
to draw substantial conclusions.  The brightness and complexity of the
feature make \ugc\ a worthy target for follow-up observations with
\xmm, \chandra, \astroe, and in the future {\it XEUS}.


\section*{Acknowledgements}

Based on observations obtained with \xmm, an ESA science mission with
instruments and contributions directly funded by ESA Member States and
the USA (NASA). Many thanks to Frank Haberl, Michael Freyberg, Wolfgang
Brinkmann, and G\"unther Hasinger for helpful discussions.
Much appreciation to the anonymous referee for providing a very helpful 
report.



\bsp
\label{lastpage}
\end{document}